\documentclass[reprint,aps,twocolumn]{revtex4-1}
\usepackage{amsmath}
\usepackage{amssymb}
\usepackage{graphicx}
\usepackage{subfigure}
\usepackage{algorithm}
\usepackage{bm}
\usepackage{color}

\usepackage{commath}
\usepackage{dcolumn}
\usepackage{multirow}

\usepackage{hyperref}

\begin{document}

\title{$K$-core attack, equilibrium $K$-core,  and kinetically constrained spin system}

\author{Hai-Jun Zhou$^{1,2,3}$}

\affiliation{$^1$Institute of Theoretical Physics, Chinese Academy of Sciences, Zhong-Guan-Cun East Road 55, Beijing 100190, China}

\affiliation{$^2$MinJiang Collaborative Center for Theoretical Physics, MinJiang University, Fuzhou 350108, China}

\affiliation{$^3$School of Physical Sciences, University of Chinese Academy of Sciences, Beijing 100049, China}

\date{\today}

\begin{abstract}
  Kinetically constrained spin systems are toy models of supercooled liquids and amorphous solids. In this \emph{Perspective}, we revisit the prototypical Fredrickson-Andersen (FA) kinetically constrained model from the viewpoint of $K$-core combinatorial optimization. Each kinetic cluster of the FA system, containing all the mutually visitable microscopic occupation configurations, is exactly the solution space of a specific instance of the $K$-core attack problem. The whole set of different jammed occupation patterns of the FA system is the configuration space of an equilibrium $K$-core problem. Based on recent theoretical results achieved on the $K$-core attack and equilibrium $K$-core problems, we discuss the thermodynamic spin glass phase transitions and the maximum occupation density of the fully unfrozen FA kinetic cluster, and the minimum occupation density and extreme vulnerability of the partially frozen (jammed) kinetic clusters. The equivalence between $K$-core attack and the fully unfrozen FA kinetic cluster also implies a new way of sampling $K$-core attack solutions.
\end{abstract}

\maketitle

\section{Introduction}
\label{sec:Intro}

A graph $G$ is formed by $N$ vertices and the edges linking between these vertices. Two vertices $i$ and $j$ are said to be nearest neighbors if they are connected by an edge, which is often denoted as $(i, j)$. The degree $d_i$ of vertex $i$ is defined as the number of attached edges. The mean vertex degree is the averaged value of $d_i$ over all the vertices, and it is denoted as $D$ in the present paper. Lattice structures are typical graphs encountered in the field of condensed-matter physics, finite-connectivity random graphs are widely adopted in theoretical computer science and applied mathematics, and small-world and scale-free complex networks are common in real-life social and technological systems~\cite{He-Liu-Wang-2009,Li-etal-2008}.

An important collective and global structural property of a graph is its $K$-core~\cite{Kong-etal-2019,DeGregorio-etal-2016}. According to the strict definition, the $K$-core of a graph $G$ is the unique subgraph induced by a maximal set of vertices under the following simple bootstrap criterion: a vertex $i$ is a member of the $K$-core if and only if at least $K$ of its $d_i$ nearest neighbors are also members of the $K$-core~\cite{Chalupa-Leath-Reich-1979,Seidman-1983}. As a simple positive example, we see that if $d_i \geq K$ for all the vertices, the whole graph $G$ is a $K$-core. Given a graph, we can get its $K$-core by recursively deleting the vertices that are connected to fewer than $K$ other vertices. If a non-empty subgraph survives this pruning process, then it is the $K$-core of the original graph, otherwise the graph has no $K$-core. For example, if a graph is a collection of tree components and there are no loops within the graph, then it contains no $K$-core, for any $K\geq 2$.

We can generalize the concept of $K$-core to incorporate the more common situations of different vertices having different threshold values. Given a set $\{ k_i: i \in G\}$ of $N$ non-negative integers, we can define the $\{k_i\}$--core as the unique maximal subgraph with each member vertex $i$ being accompanied by at least $k_i$ nearest neighboring member vertices~\cite{Branco-1993}. The $K$-core is simply the special case of $\{k_i\}$--core with $k_i = K$ for all the vertices. To be concise, we will assume $k_i = K$ in the following discussions. The extension to the more general $\{k_i\}$--core is straightforward.

There have been many inspiring examples of applying $K$-core to various structural and dynamical issues of complex networks science, see Ref.~\cite{Kong-etal-2019} for a comprehensive review. The present \emph{Perspective} is much limited in scope. We will only review some recent theoretical work done on two $K$-core combinatorial optimization problems. The emphasize is on the deep relevance of these $K$-core spin glass problems to the statistical mechanics of kinetically constrained spin models. Especially we will make explicit and obvious the intrinsic link between the $K$-core attack problem and the configuration space of kinetically constrained spin models (Sec.~\ref{sec:global}). The methodological implications of this bi-directional link will be discussed. 

Kinetically constrained spin systems are simple and popular graphical toy models for understanding the glassy dynamics of supercooled liquids and the jammed packing of amorphous solids~\cite{Ritort-Sollich-2003,Garrahan-etal-2011,Jack-2020}. The Fredrickson-Andersen (FA) model is a prototypical and the simplest kinetically constrained spin system~\cite{Fredrickson-Andersen-1984}. It assigns a binary occupation state, empty or being occupied, to every vertex of the graph $G$. For a vertex $i$ to flip its state, however, it must have less than $k_i$ occupied nearest neighbors at the moment of flipping. The FA model and other kinetically constrained spin models have served as basic theoretical platforms to study the dynamical phase transitions of supercooled liquids~\cite{Ritort-Sollich-2003,Garrahan-etal-2011,Tang-etal-2024,Franz-etal-2016,Foini-etal-2012}. For simplicity, we will consider only the FA model here (Sec.~\ref{sec:FA}).

The local kinetic rule of the FA system leads to kinetic ergodicity-breaking in the configuration space and the emergence of many kinetic clusters, each of which contains all the mutually visitable microscopic occupation configurations (Sec.~\ref{sec:keb}). The one-to-one mapping between kinetic clusters and specific instances of the $K$-core attack problem (Sec.~\ref{sec:global}) implies a principled way of studying the statistical properties of individual kinetic clusters. We focus on the $K$-core attack problem in Section~\ref{sec:kct}, and discuss the issue of thermodynamic phase transitions and maximum unjammed packing. It will become clear that the spin glass transitions in single FA kinetic clusters are induced not by the formation of $K$-core, but by the global constraint of $K$-core \emph{absence}.

The local kinetic rule of the FA system is, at least apparently, identical to the local $K$-core static constraint. Because of this similarity, the dynamical properties of the FA kinetic system has been studied from the perspective $K$-core percolation~\cite{Adler-1991,Rizzo-2018,Perrupato-Rizzo-2022,Perrupato-Rizzo-2023}, which manifests as the sudden emergence of an extensive $K$-core of occupied vertices as the fraction of occupied vertices increases beyond certain threshold value $p_c$~\cite{Branco-1993,Pittel-etal-1996,Dorogovtsev-etal-2006,Zhao-Zhou-Liu-2013,Cellai-etal-2011,Baxter-etal-2011}. This discontinuous $K$-core phase transition (also referred to as the bootstrap transition) has been associated with the jamming transition of structural glasses~\cite{Sellitto-etal-2005,Schwarz-Liu-Chayes-2006,Morone-etal-2019}. We will briefly review the $K$-core phase transition in Sec.~\ref{sec:kpl}. Following this, we will then review some recent theoretical efforts on the equilibrium $K$-core problem, and discuss its relevance with the jammed kinetic clusters of the FA system (Sec.~\ref{sec:kfr}). Results on minimum jammed packing and on extreme vulnerability of jamming will be demonstrated.

All the theoretical results discussed in the present paper were obtained on random graph ensembles. To deeply appreciate the thermodynamic properties of finite-dimensional FA kinetic systems, we need to extend the investigations on $K$-core attack and equilibrium $K$-core optimization problems to finite-dimensional lattices (outlook in Sec.~\ref{sec:conc}). Hopefully, this \emph{Perspective} might be useful and stimulating at least to some researchers working on kinetic constrained models or on $K$-core optimization problems.

\section{The Fredrickson-Andersen model}
\label{sec:FA}

The Fredrickson-Andersen model was a prototypical kinetically constrained spin system introduced in the early 1980s for the purpose of understanding the dynamical properties of supercooled liquids and structural glasses~\cite{Fredrickson-Andersen-1984}. It has inspired many other kinetically constrained spin models and has stimulated a lot of theoretical and simulation studies~\cite{Ritort-Sollich-2003,Garrahan-etal-2011,Jack-2020}. The FA model is defined on a finite-connectivity graph $G$ of $N$ vertices~\cite{Fredrickson-Andersen-1984}. Each vertex with index $i \in \{1, 2, \ldots, N\}$ has a binary occupation state, $c_i = 0$ (vertex being empty) or $c_i = 1$ (vertex being occupied). A microscopic configuration of the whole system at any time is the collection of all these individual states, $\vec{\bm{c}} \equiv (c_1, c_2, \ldots, c_N)$, and the total number of possible configurations is $2^N$. All these configurations are allowed for the system. This is quite different from the static rule of the lattice glass models that were introduced in the early 2000s~\cite{Biroli-Mezard-2002,Rivoire-etal-2004}. In a lattice glass, an occupied vertex prohibits its nearest neighbors from being too densely occupied. Such local and static excluded-volume effects are completely absent in the FA system.

The total energy $E$ of a microscopic configuration $\vec{\bm{c}}$ of the FA system is defined as
\begin{equation}
  E( \vec{\bm{c}} ) = - \sum\limits_{i=1}^N c_i \; .
  \label{eq:FAenergy}
\end{equation}
The energy contribution ($-c_i$) of each vertex $i$ depends only on its occupation state.  The occupied state $c_i = 1$ is favored in the energetic sense, and the fully occupied configuration $(1, 1, \ldots, 1)$ is the unique minimum-energy configuration of the system. Because there is no interaction energy between any pair of neighboring vertices, the equilibrium Boltzmann distribution $P_{\textrm{B}}(\vec{\bm{c}})$ of model (\ref{eq:FAenergy}) is factorized,
\begin{equation}
  P_{\textrm{B}}( \vec{\bm{c}} ) =  \prod\limits_{i=1}^{N} q(c_i ) \; ,
  \quad \textrm{with}\ \  
  q(c_i ) = \frac{e^{\beta c_i}}{1 + e^{\beta}} \; .
  \label{eq:Boltzmann}
\end{equation}
Here $\beta$ is the inverse temperature of the environment, and $q(c_i)$ is the marginal probability of the binary occupation state $c_i$.

We follow this factorized probability distribution to sample an initial configuration (denoted as $\vec{\bm{c}}^{\,0}$) at certain inverse temperature $\beta_0$. The initial states of the $N$ vertices are independent, and every vertex is occupied with the same probability. The mean fraction $p$ of initially occupied vertices is then
\begin{equation}
  p \ = \ \frac{1}{1 + e^{-\beta_0}} \; .
\end{equation}

Starting from such an initial configuration $\vec{\bm{c}}^{\,0}$ with occupation density $p$, the system then evolves with time and finally reaches equilibrium at another inverse temperature $\beta$. The inverse temperature $\beta$ can be the same as the value $\beta_0$ of sampling $\vec{\bm{c}}^{\,0}$. It can also be different from $\beta_0$, to allow for cooling or heating in the system and for changes in the mean occupation density.

The FA system explores the configuration space by single-vertex state flips. At each elementary time step (say $\Delta t = 1/N$), a vertex $i$ is picked up uniformly at random from the $N$ vertices, and its state is flipped to the opposite one with probability $w_{c_i \rightarrow 1-c_i}$  and kept unchanged with the remaining probability $1 - w_{c_i\rightarrow 1-c_i}$. The flipping probability strongly depends on the states of the nearest neighbors of this vertex $i$. Denote $\partial i \equiv \{j : (i, j) \in G\}$ as the vertex subset containing all the nearest neighbors. If there are $k_i$ or more occupied nearest neighbors and
\begin{equation}
  \sum\limits_{j\in \partial i} c_j \ \geq \ k_i \; ,
  \label{eq:kc1}
\end{equation}
then vertex $i$ is blocked from changing state, and $w_{c_i \rightarrow 1-c_i} = 0$ independent of the particular value of $c_i$.

The empty state has a facilitation effect. In the opposite (facilitated) situation of fewer than $k_i$ occupied nearest neighbors,
\begin{equation}
  \sum\limits_{j\in \partial i} c_j \ <  \ k_i \; ,
  \label{eq:kc2}
\end{equation}
vertex $i$ is free to change state. The flipping probabilities $w_{1\rightarrow 0}$ and $w_{0\rightarrow 1}$ then satisfy the detailed balance condition at inverse temperature $\beta$,
\begin{equation}
  \frac{w_{0\rightarrow 1}}{w_{1\rightarrow 0}} = e^{\beta} \; .
\end{equation}
A convenient choice is to set
\begin{equation}
  w_{0\rightarrow 1} = \frac{1}{1 + e^{-\beta}}
  \; ,
  \quad
  w_{1\rightarrow 0} = \frac{1}{1+e^{\beta}}  
  \; .
\end{equation}

Notice that under the condition of fixed neighbor states, the state flip between $c_i=0$ and $c_i=1$ for each vertex $i$ is locally reversible. The local equilibrium distribution of $c_i$ under such a facilitated boundary condition is described by the same expression $q(c_i)$ of Eq.~(\ref{eq:Boltzmann}).

For simplicity, we will assume all the threshold numbers $k_i$ in Eqs.~(\ref{eq:kc1}) and (\ref{eq:kc2}) are identical to the same integer value $K$, with $K \geq 2$.

\section{Kinetic ergodicity-breaking}
\label{sec:keb}

Because the state of any vertex $i$ is not allowed to flip under the kinetic blocking condition (\ref{eq:kc1}), the configuration space of the FA system is not kinetically ergodic as a whole. The $2^N$ microscopic configurations are distributed to different kinetic clusters $C_\alpha$ with indices $\alpha \in\{1, \ldots, \mathcal{N}\}$, where $\mathcal{N}$ denotes the total number of kinetic clusters. This is a phenomenon of kinetic ergodicity-breaking.

\subsection{The fully unfrozen kinetic cluster $C_1$}

Let us assign the cluster index $\alpha = 1$ to the kinetic cluster which contains the fully empty configuration $\vec{\bm{0}}$ with all vertices being empty. We put to this cluster $C_1$ all the other configurations that are visitable from this seed configuration. Since each state flip is locally reversible, if a configuration $\vec{\bm{c}}^{\,\prime}$ can be visited from another configuration $\vec{\bm{c}}$ through a sequence of allowed single-vertex flips, the reverse is also true: configuration $\vec{\bm{c}}$ can be visited from $\vec{\bm{c}}^{\,\prime}$ through a sequence of allowed single-vertex flips. The cluster $C_1$ therefore contains configuration $\vec{\bm{0}}$ and all the other configurations that can reach and be reached from $\vec{\bm{0}}$ through a set of allowed single-vertex flips.

For each vertex $i$, there is at least one configuration $\vec{\bm{c}} \in C_1$ in which its state $c_i = 1$ (e.g., the one with all the other vertices being empty) and there is at least one configuration in which its state $c_i = 0$. In other words, every vertex is unfrozen in the kinetic cluster $C_1$.

Notice that if a vertex $i$ is unfrozen in certain kinetic cluster $C_\alpha$ (not necessarily $C_1$), and its state is $c_i = 1$ in one configuration $\vec{\bm{c}}$ of this cluster, then the other configuration $\vec{\bm{c}}^{\,\prime}$ which differs with $\vec{\bm{c}}$ only at vertex $i$ ($c_i = 0$) must also belong to the same kinetic cluster $C_\alpha$. We follow the elegant arguments of Perrupato and Rizzo to prove this statement~\cite{Perrupato-Rizzo-2023}.

Since vertex $i$ is unfrozen, there must be a sequence of allowed single-vertex flips, $\vec{\bm{c}} \rightarrow \vec{\bm{c}}^{\,1} \rightarrow \ldots \rightarrow \vec{\bm{c}}^{\,n}$, with the last configuration $\vec{\bm{c}}^{\,n}$ having $c_i = 0$. Then the reverse of this trajectory but with $c_i$ being fixed to $c_i = 0$ must also be an allowed kinetic trajectory. This modified reverse trajectory leads to the configuration $\vec{\bm{c}}^{\,\prime}$ which is identical to $\vec{\bm{c}}$ except for $c_i$. This means that $\vec{\bm{c}}$ and $\vec{\bm{c}}^{\,\prime}$, which differ only at vertex $i$, are mutually visitable and are members of the same cluster $C_\alpha$.

\begin{figure}[b]
  \centering
  \includegraphics[width=0.7\linewidth]{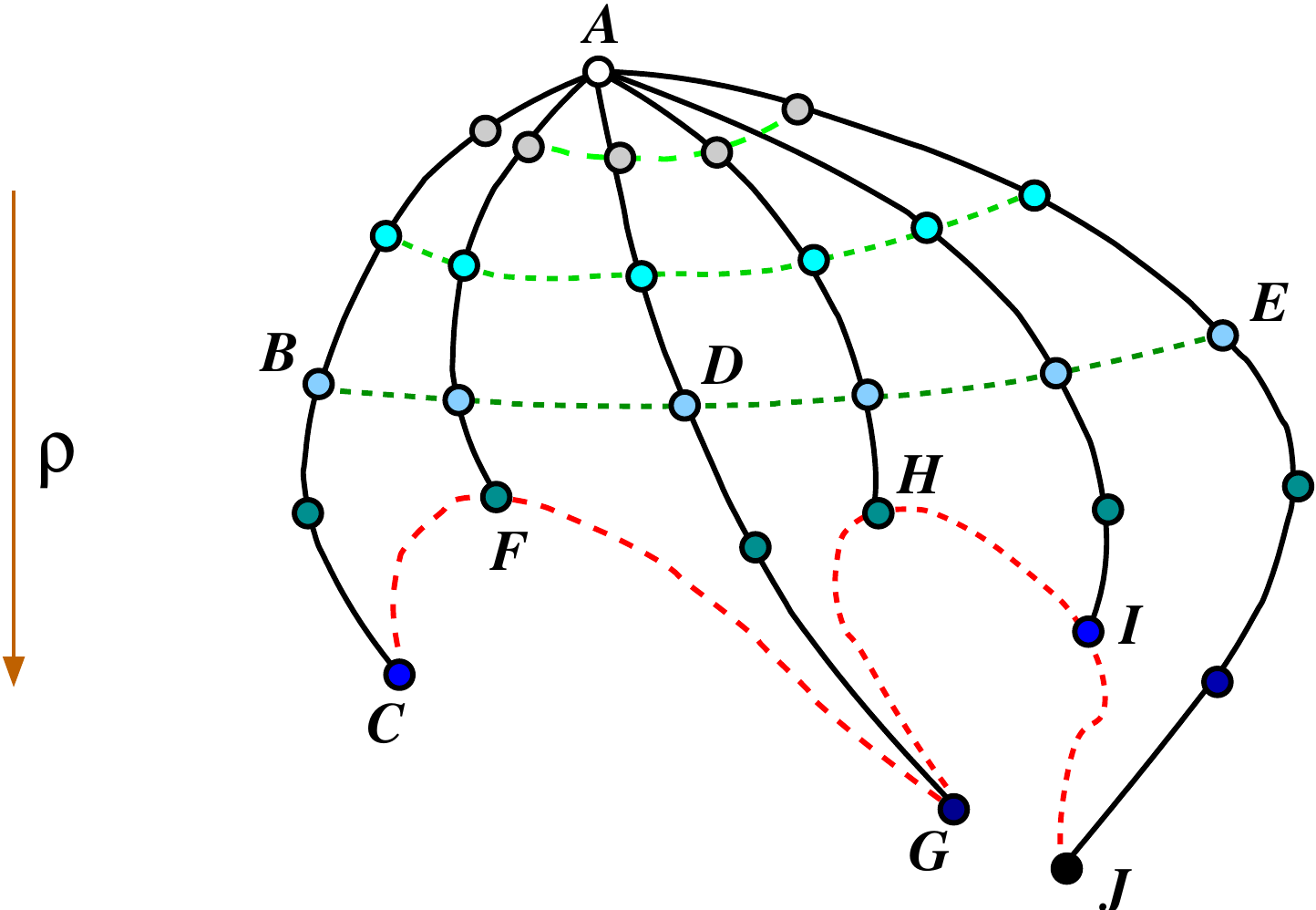}
  \caption{
    Schematic structure of the kinetic cluster $C_1$. Circles denote individual configurations $\vec{\bm{c}}$, partially ranked according to the fraction $\rho$ of occupied vertices. Point $A$ on the top represents the fully empty configuration $\vec{\bm{0}}$. Points $C$, $G$ are  two local maximally occupied configurations, and point $J$ denotes a global maximal occupied configuration. Lines linking the circles indicate single-vertex flipping trajectories. The path $B$-$D$-$E$ is a trajectory with $\rho$ being approximately constant. The curved trajectories $C$-$F$-$G$ and $G$-$H$-$I$-$J$ indicate that the transitions between configurations $C$ and $G$ and between $G$ and $J$ must overcome large energy barriers.
  }
  \label{fig:C1}
\end{figure}

Because of this co-appearance property, we can prove that $C_1$ is the only kinetic cluster in which all the vertices are unfrozen. Suppose this statement is false and there is another kinetic cluster (say $C_\alpha$) in which all the vertices are also unfrozen. Then if $\vec{\bm{c}} \in C_\alpha$ is a configuration with $c_i = 1$, the other configuration $\vec{\bm{c}}^{\,\prime}$ which is identical to $\vec{\bm{c}}$ except for vertex $i$ must also belong to $C_\alpha$. If some vertices (say $j$) are occupied in configuration $\vec{\bm{c}}^{\,\prime}$, we can again flip $c_j$ to $c_j = 0$ to get another configuration $\vec{\bm{c}}^{\,\prime\prime}$ that is reachable from $\vec{\bm{c}}^{\,\prime}$. Continue this flipping process, we will finally arrive at the all-empty configuration $\vec{\bm{0}}$, which belongs to kinetic cluster $C_1$. Therefore the fully unfrozen cluster $C_\alpha$ must be identical to $C_1$.

To complement these above discussions, we draw in Fig.~\ref{fig:C1} a schematic picture for the organization of microscopic configurations in the fully unfrozen cluster $C_1$.

\subsection{Fully frozen and partially frozen clusters}
\label{subsec:Calpha}

At the other limit is the cluster $C_{\mathcal{N}}$ with the maximum index $\alpha = \mathcal{N}$. It contains a single configuration $\vec{\bm{1}}$ with all the vertices being occupied. Every vertex is obviously frozen in this single-configuration cluster.

There are other frozen clusters with indices $\alpha < \mathcal{N}$. If every empty vertex of a configuration $\vec{\bm{c}}$ has $\geq K$ occupied nearest neighbors, and every occupied vertex of this configuration also has this property, then $\vec{\bm{c}}$ is a fully frozen configuration and it forms a fully frozen kinetic cluster. It is quite easy to write down many such fully frozen configurations for a given input graph $G$.

Besides fully unfrozen and fully frozen kinetic clusters, there are other partially frozen kinetic clusters $C_\alpha$, each of which contains at least two (but often many) microscopic configurations. Some of the vertices are frozen in such a cluster, while the states of the remaining vertices are unfrozen. 

A kinetic cluster $C_\alpha$ can be fully specified by its set ($F_1$) of frozen occupied vertices and the set ($F_0$) of frozen empty vertices. If two kinetic clusters have identical set $F_1$ and identical set $F_0$, they must be the same cluster. To prove this, one only needs to show that the configuration with all the unfrozen vertices being in the empty state must belong to both clusters, and hence these two clusters must be the same.

\section{Global constraint associated with a single kinetic cluster}
\label{sec:global}

Each single kinetic cluster $C_\alpha$ contains a maximal set of microscopic configurations $\vec{\bm{c}}$ that are mutually visitable through a sequence of allowed single-vertex flips. There may be strong correlations among the occupation states of the unfrozen vertices of this cluster $C_\alpha$. We now prove a simple but highly nontrivial fact, that all these possibly complicated correlations can be encoded in a single global constraint.

Let us first consider the fully unfrozen cluster $C_1$, which contains all the configurations that are visitable from the fully-empty $\vec{\bm{0}}$. Every vertex $i$ can take both states $c_i = 0$ and $c_i = 1$. Given a microscopic configuration $\vec{\bm{c}} \in C_1$, let us denote by $U_0$ the set of empty vertices:
\begin{equation}
  U_0( \vec{\bm{c}} ) \ =  \ \{ i : c_i = 0 \} \; .
  \label{eq:Gamma0u}
\end{equation}
Since there are no frozen vertices in $C_1$, all the other vertices not belonging to $U_0$ are in the occupied state. It is obvious that the subgraph induced by these occupied vertices of $\vec{\bm{c}}$ should not form a $K$-core, because otherwise some of these vertices will be permanently frozen.

The reverse statement is also true. If the subgraph formed by the occupied vertices of a configuration $\vec{\bm{c}}$ and the edges between them is not sustaining a $K$-core, this configuration must belong to the fully unfrozen cluster $C_1$, because there is a sequence of allowed single-vertex flips linking configuration $\vec{\bm{c}}$ to the fully-empty $\vec{\bm{0}}$.

The above two statements mean that the set $U_0( \vec{\bm{c}})$ of empty vertices of each configuration $\vec{\bm{c}} \in C_1$ must be a $K$-core attack set: The residual subgraph of $G$ after deleting all the vertices of set $U_0(\vec{\bm{c}})$ contains no $K$-core~\cite{Zhou-2022}. And if $U_0( \vec{\bm{c}})$ is a $K$-core attack set, then the configuration $\vec{\bm{c}}$ must be a member of the fully unfrozen cluster $C_1$.

Therefore, the configuration subspace $C_1$ is equivalent to all the $K$-core attack sets $U_0$ of the graph $G$. The statistical physics property of the fully unfrozen cluster $C_1$ can then be explored by studying the solution space of the corresponding $K$-core attack problem. This idea will be followed in the next section~\ref{sec:kct}.

We make an interesting remark here. This significant link between the configuration subspace $C_1$ of the FA system and the $K$-core attack problem also implies a new simulation algorithm to sample solutions for the $K$-core attack problem. We may adopt the single-vertex flip dynamics of the FA system to obtain equilibrium $K$-core attack solutions, without checking the global $K$-core-absence constraint. We will explore this idea in a separate paper, here we focus on the implications of the $K$-core attack problem to the FA system. (For this sampling purpose, to accelerate the Markov-chain dynamics, we can also add the elementary process of swapping the states of two flippable vertices $i$ and $j$, without violating the FA kinetic rule.) 

The above-established equivalence relationship also extends to the partially frozen kinetic clusters $C_\alpha$. For such a partially frozen cluster, we also denote the set of empty unfrozen vertices of each configuration $\vec{\bm{c}} \in C_\alpha$ as $U_0( \vec{\bm{c}})$, namely
\begin{equation}
  U_0\bigl( \vec{\bm{c}} \bigr) 
  \ = \
  \{i : c_i = 0\; ,  \textrm{vertex}\; i \; \textrm{being unfrozen}
  \} \; .
\end{equation}
Obviously, all the occupied unfrozen vertices of the configuration $\vec{\bm{c}}$ should not belong to the $K$-core. Therefore this set $U_0$ of empty vertices must be a $K$-core attack set for all the unfrozen vertices of the kinetic cluster $C_\alpha$.

\section{Exploring the fully unfrozen kinetic cluster}
\label{sec:kct}

The discussion of the preceding section~\ref{sec:global} convinces us that each kinetic cluster $C_\alpha$ of the FA model contains all the solutions of a spin glass model with a global constraint of $K$-core absence. This perfect superposition of configuration spaces implies that, to explore and appreciate the rich statistical property of individual kinetic clusters, we could turn the kinetic FA system into a static spin glass model and then treat this static model by the advanced methods of equilibrium statistical mechanics.  

We begin by defining the $K$-core attack problem more precisely. To be concrete, let us focus on $C_1$, the kinetic cluster free of any frozen vertices. Each partially frozen kinetic cluster $C_\alpha$ can be treated in the same way. The only slight complication is that the spin glass model should be defined only on the set of unfrozen vertices, with some environmental pinning effects induced by the frozen vertices.

\subsection{The $K$-core attack problem}

As mentioned earlier, the $K$-core of a graph $G$ is the unique maximal subgraph in which every member vertex is connected to at least $K$ other member vertices. This unique $K$-core can be determined through a simple vertex pruning process. As long as there is a vertex whose degree is less than $K$, it is deleted from the graph together with its attached edges. If a non-vanishing subgraph survives this iterative pruning, it is the $K$-core.

Given an input graph $G$ of $N$ vertices, let us assign a binary occupation state $c_i$ to each vertex $i$, $c_i = 1$ (occupied) or $c_i = 0$ (empty). The total number of possible microscopic configurations is simply $2^N$. If a vertex $i$ is occupied, its energy is $E_i = -1$, otherwise its energy is $E_i = 0$. We impose a global constraint on the occupation configuration as follows: The subgraph induced by the occupied vertices should not contain a $K$-core. This means that if we delete all the empty vertices from the graph $G$, the remaining subgraph should completely dissolve under the $K$-core pruning process.

Under this $K$-core-absence global constraint, we define an equilibrium partition function $Z^{\textrm{atk}}(\beta)$ for our $K$-core attack problem as
\begin{equation}
  Z^{\textrm{atk}}(\beta) = \sum\limits_{c_1, \ldots, c_N}  I_{0}( c_1, \ldots, c_N)
  \prod\limits_{i=1}^{N} e^{\beta c_i}
  \; .
  \label{eq:Zattack}
\end{equation}
Here $I_0(c_1, \ldots, c_N) \in \{0, 1\}$ is the $K$-core-absence indicator function: $I_{0}= 1$ if the occupied vertices of the configuration $\{c_1, \ldots, c_N\}$  \emph{does not} sustain a $K$-core, otherwise $I_{0} = 0$. This partition function is a weighted summation over all the configurations of the unfrozen kinetic cluster $C_1$. If a satisfying configuration has $N_1$ occupied vertices then its contribution is $e^{\beta N_1}$.

The $K$-core attack problem is a challenging combinatorial optimization problem~\cite{Guggiola-Semerjian-2015,Zhao-etal-2015,Yuan-etal-2016,Schmidt-etal-2018,Zhou-2022,Zhou-Zhou-2023}. A special case is the $2$-core attack problem ($K=2$), which means destroying all the loops in the graph by removing a set of vertices, and it is also known as the feedback vertex set problem~\cite{Fomin-etal-2008,Zhou-2013}. In the literature, the practical significance of the feedback vertex set problem has been linked to identifying influential spreaders in a complex network~\cite{Morone-Makse-2015,Mugisha-Zhou-2016,Braunstein-etal-2016,Zdeborova-etal-2016,Chujyo-Hayashi-2021}, see also review articles~\cite{Lv-etal-2016,Pei-etal-2019}.

At this point, it is helpful to emphasize again the fundamental difference between the $K$-core attack model (\ref{eq:Zattack}) for the kinetically constrained FA system and the lattice glass model~\cite{Biroli-Mezard-2002,Rivoire-etal-2004}. The latter is characterized by local excluded-volume constraints, such that each occupied vertex $i$ only allows fewer than $k_i$ of its nearest neighbors to be occupied. Some well-known nondeterministic polynomial-complete (NP-complete) optimization problems, such as the vertex cover problem~\cite{Weigt-Hartmann-2001,Zhou-2003a,Zhao-Zhou-2014} and the dominating set problem~\cite{Haynes-Hedetniemi-Slater-1998,Echenique-etal-2005,Zhao-Habibulla-Zhou-2015}, can also be viewed as lattice glass models. It is now well accepted that these local geometric constraints can lead to ergodicity-breaking in the thermodynamic sense (see also related work linking to the $p$-spin interaction models~\cite{Franz-etal-2016,Foini-etal-2012}). Various spin glass macroscopic states will emerge at high densities of occupied vertices.

In contrast, there are no local constraints in the NP-complete $K$-core attack model (\ref{eq:Zattack}). All the $2^{1+d_i}$ possible local occupation arrangements concerning a vertex $i$ and its $d_i$ nearest neighbors are allowed. The expected low-temperature (high-$\beta$) glass phases of this model are completely caused by a single global constraint, that the occupied vertices should not build a $K$-core. As the $K$-core-absence constraint is a direct consequence of kinetic visitability within cluster $C_1$, the low-temperature glass phases of the FA kinetic system, if exist, are essentially thermodynamic phases induced by local kinetic rules.

Stimulated by the breakthrough made on the random $K$-satisfiability problem in $2002$~\cite{Mezard-etal-2002}, many NP-complete combinatorial optimization problems have been studied by the mean field cavity method of statistical physics during the last two decades~\cite{Mezard-Parisi-2001,Mezard-Parisi-2003,Krzakala-etal-PNAS-2007,Mezard-Montanari-2009,Zhou-2015}. We can also study the $K$-core attack model (\ref{eq:Zattack}) by the spin glass cavity method, but a main technical complication is that the indicator function $I_0(c_1, \ldots, c_N)$ depends on the states of all the $N$ vertices. We need to find an efficient way to implement the $K$-core-absence global constraint into a set of local constraints.

There are two different strategies in the literature. The direct and simple strategy is to assign to each vertex $i$ an integer time label $t_i \geq 0$~\cite{Altarelli-Braunstein-DallAsta-Zecchina-2013,Altarelli-Braunstein-DallAsta-Zecchina-2013b,Guggiola-Semerjian-2015}. The initially deleted vertices all have time label $t=0$ and all the other vertices have positive time labels. The local constraint associated with a vertex $i$ can be constructed based on the $K$-core pruning dynamics. This implementation is rigorous, but each vertex has many different states and the configuration space of the resulting model is very huge. The other strategy is to map the $K$-core attack problem to a polymer packing problem~\cite{Zhou-2013,Li-etal-2021,Zhou-2022,Zhou-Zhou-2023}. Each polymer is essentially in the shape of a directed tree, with the direction marking the partial time order of the $K$-core pruning dynamics. Many such polymers are packed together in graph $G$ under some local constraints. This second implementation is not completely equivalent to the $K$-core-absence global constraint. Its advantage is that each vertex only has a small number of states. The message-passing algorithms inspired by this implementation are very efficient and are able to construct close-to-minimum $K$-core attack solutions for single random graph instances~\cite{Zhou-2022,Zhou-Zhou-2023}. 

Without delve into more technical details, we now review in the following two subsections some main theoretical results obtained on random graphs. Unlike lattice systems, there are no structure in a random graph $G$. All the edges in such a graph are randomly connected, without any constraint or with the local constraints of fixed vertex degrees. Such a random graph is locally tree-like, and the typical length of loops diverges logarithmically with graph size $N$~\cite{He-Liu-Wang-2009,Li-etal-2008}.

\subsection{Entropy and maximum occupation density}

The partition function (\ref{eq:Zattack}) contains rich information about the energy landscape of the configuration subspace $C_1$. The mean occupation density $\rho$ of the graph is defined as the average of the number $N_1$ (rescaled by $N$) of occupied vertices among all the configurations that satisfy $I_0 = 1$, each with weight $e^{\beta N_1}$. The abundance of microscopic configurations at each occupation density $\rho$ is measured by the entropy density $s(\rho)$, which can be deduced from the thermodynamic relationship $Z^{\textrm{atk}}(\beta) \approx e^{\beta N \rho} e^{N s}$, namely
\begin{equation}
  s = -  \beta \rho +  \frac{1}{N} \ln \Bigl[ Z^{\textrm{atk}}(\beta) \Bigr] \; .
\end{equation}
\begin{figure}[b]
  \centering
  \includegraphics[angle=270,width=0.7\linewidth]{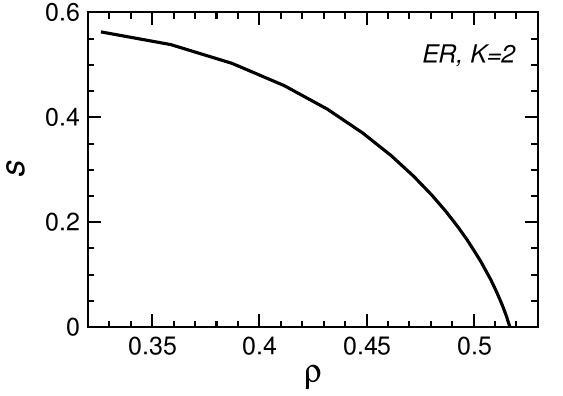}
  \caption{
    Entropy density $s(\rho)$ of the $2$-core attack problem for the Erd\"os-R\'enyi (ER) random graph ensemble of mean vertex degree $D=10.0$, as a function of the density $\rho$ of occupied vertices. Results are obtained by the replica-symmetric mean field theory~\cite{Zhou-2013}.
  }
  \label{fig:ERc10s}
\end{figure}

The mean occupation density $\rho$ increases with the inverse temperature $\beta$. When $\rho$ becomes sufficiently large, the entropy density $s(\rho)$ as a function of $\rho$ shows a decreasing trend, and it decreases to zero as $\rho$ approaches certain value $\rho_{\textrm{max}}$ from below. This threshold value $\rho_{\textrm{max}}$ is the maximum occupation (packing) density that is achievable in the kinetic cluster $C_1$, and the corresponding configurations are the densest packing configurations of this kinetic cluster $C_1$.

As an example, we show in Fig.~\ref{fig:ERc10s} the theoretical entropy density curve $s(\rho)$ for the $2$-core attack problem ($K=2$) on the ensemble of maximally random graph of mean vertex degree $\langle d_i \rangle = D = 10.0$~\cite{Zhou-2013}. Such random graphs are referred to as Erd\"os-R\'enyi graphs in the literature~\cite{He-Liu-Wang-2009,Li-etal-2008}. The $(D/2) N$ edges in such a graph are randomly and independently drawn from the $N(N-1)/2$ candidate edges, so the vertex degrees are not exactly the same but are distributed around the mean value $D$.

We see that $s(\rho)$ is a concave function of occupation density $\rho$. The entropy density approaches zero when $\rho$ reaches the maximum value $\rho_{\text{max}} = 0.517$. This means that the densest packing as kinetically achievable from the fully empty configuration can not exceed the density $0.517$. Numerical results obtained by two different heuristic optimization algorithms on single graph instances have confirmed this prediction~\cite{Zhou-2013,Qin-Zhou-2014}. This example demonstrates that the static model (\ref{eq:Zattack}) makes it possible to compute the maximum packing density of the fully unfrozen kinetic cluster $C_1$. We also notice that, for $K = 2$, the maximum packing density $\rho_{\textrm{max}}$ on regular random graph ensembles of degree $D$ has been rigorously upper-bounded~\cite{Bau-Wormald-Zhou-2002}.

\begin{table}[t]
    \caption{
      Algorithmic results on the maximum density $\rho_{\textrm{max}}$ of occupied vertices for the $K$-core attack problem, obtained by the {\tt hCTGA}~\cite{Zhou-Zhou-2023} and the {\tt WN}~\cite{Schmidt-etal-2018} algorithms on regular random graphs of degree $D$ and size $N=10000$.  The last two columns list the mean-field (MF) predictions based on two different models~\cite{Guggiola-Semerjian-2015,Zhou-2013}.
    }
    \label{tab:RR}

    \centering
    \begin{tabular}{cccccc}
      \\
      \hline\hline
      $D$ \  & $K$ \  & {\tt hCTGA}~\cite{Zhou-Zhou-2023} \
      & {\tt WN}~\cite{Schmidt-etal-2018} \
      & MF~\cite{Guggiola-Semerjian-2015} \
      & MF~\cite{Zhou-2013} \
      \\
      \hline
      3 & 2 & 0.7499 & 0.7497  & 0.7500 & 0.7500 \\
      \hline
      4 & 2 & 0.6665 & 0.6622  & 0.6667 & 0.6667 \\
      & 3 & 0.9282 & 0.9253  & 0.9537 & \\
      \hline
      5 & 2 & 0.6207 & 0.6036 & 0.6215 & 0.6216 \\
      & 3 & 0.8123 & 0.8122 & 0.8333 & \\
      & 4 & 0.9736 & 0.9719 & 0.9869 & \\
      \hline
      6 & 2 & 0.5748 & 0.5556  & 0.5774 & 0.5770 \\
      & 3 & 0.7379 & 0.7351  & 0.7500 & \\
      & 4 & 0.8925 & 0.8915  & 0.9238 & \\
      & 5 & 0.9872 & 0.9862  & 0.9943 & \\
      \hline
      7 & 2 & 0.5359 & 0.5164  & 0.5400 & 0.5385 \\
      & 3 & 0.6885 & 0.6791  & 0.7000 & \\
      & 4 & 0.8199 & 0.8183  & 0.8500 & \\
      & 5 & 0.9309 & 0.9306  & 0.9572 & \\
      & 6 & 0.9925 & 0.9920  & 0.9969 & \\
      \hline \hline
      \\
    \end{tabular}
\end{table}

The maximum packing density $\rho_{\textrm{max}}$ depends strongly on the threshold value $K$ and increases monotonically with $K$. Table~\ref{tab:RR} presents some numerical results obtained on several regular random graph ensembles to confirm this trend. Notice that all the vertices in a regular random graph have the same vertex degree $D$. The simulation results obtained through two heuristic algorithms are also shown as comparison~\cite{Zhou-Zhou-2023,Schmidt-etal-2018}. At fixed value of $D$, we find that the relative difference between algorithmic results and the theoretical prediction increases as $K$ increases to $D-1$. This is an indication that reaching the densest packing configurations within the kinetic cluster $C_1$ is an extremely hard optimization problem.

\subsection{Spin glass phase transitions}

When the inverse temperature $\beta$ becomes sufficiently high in Eq.~(\ref{eq:Zattack}), the mean density $\rho$ of occupied vertices will be bounded above by the maximum value $\rho_{\textrm{max}}$. The limiting value may be impossible to reach in numerical simulations, and instead the low-temperature (high-$\beta$) region may be in the spin glass phase. For random graph ensembles, the existence of spin glass phase transitions have been confirmed by theoretical computations.

According to the mean-field theory of spin glasses~\cite{Mezard-Montanari-2009,Zhou-2015}, a graphical model such as (\ref{eq:Zattack}) could experience a dynamical spin glass phase transition at a critical inverse temperature $\beta_d$. As $\beta$ reaches $\beta_d$ from below, the configuration space suddenly breaks up into an exponential number of subspaces, each of which is referred to as a macroscopic state, with well-defined thermodynamic quantities such as free energy density and entropy density. There are very high free-energy barriers between these macroscopic states, so that if the system is staying at one macroscopic state with $\beta > \beta_d$, it will take a diverging time to overcome the free energy barrier to reach another different macroscopic state. The critical value $\beta_d$ for random graph systems can be computed precisely through a method of tree-reconstruction~\cite{Mezard-Montanari-2006}.

\begin{figure}[b]
  \centering
  \includegraphics[angle=270,width=0.7\linewidth]{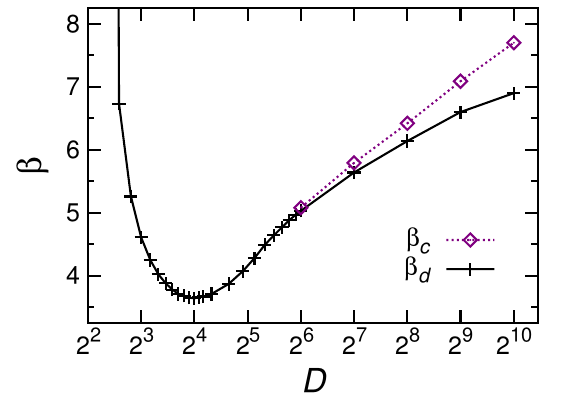}
  \caption{The inverse temperature $\beta_d$ of dynamical transition (plus symbol), and the inverse temperature $\beta_c$ of condensation transition (diamond symbol), for the $2$-core attack problem on the regular random graph ensemble with uniform vertex degree $D$~\cite{Qin-etal-2016}.
  }
  \label{fig:RRTdTk}
\end{figure}

Another important critical inverse temperature is $\beta_c$, the condensation phase transition point. At $\beta = \beta_c$, a non-exponential number of macroscopic states start to dominate the configuration space, while all the other exponential number of macroscopic states have less statistical weights compared with these dominating ones. At different values of $\beta > \beta_c$ the dominating macroscopic states may be different. For some systems the condensation transition occurs after the dynamical transition, so $\beta_c > \beta_d$. But for some other systems the two phase transitions coincide and $\beta_c = \beta_d$.  

Figure~\ref{fig:RRTdTk} shows the critical inverse temperatures $\beta_d$ and $\beta_c$, obtained for the $K$-core attack problem with $K=2$ on the regular random graph ensembles of various degrees $D$~\cite{Qin-etal-2016}. Such a phase diagram is quite similar to that of the random vertex cover problem which is a lattie glass model~\cite{Zhang-Zeng-Zhou-2009}. This similarity indicates that the thermodynamic spin glass phase transitions of the kinetic constrained model and the lattice glass model may be closely related. When $D \leq 60$ we find that $\beta_c = \beta_d$, but for higher values of $D$, we have $\beta_c > \beta_d$.

Similar spin glass phase transitions also occurs for $K \geq 3$ on random graph ensembles, but the computation of the critical values $\beta_d$ and $\beta_d$ become more difficult~\cite{Guggiola-Semerjian-2015,Zhou-2022,Zhou-Zhou-2023}. The existence of these spin glass phase transitions mean that the FA system defined on random graphs really has thermodynamic phase transitions. It is not yet clear whether these spin glass phase transitions will also occur for finite-dimensional systems. Systematic studies on the low-temperature properties of the $K$-core attack model (\ref{eq:Zattack}) on finite-dimensional systems are needed to fully resolve this fundamental issue.

For $\beta$ higher than the condensation transition value $\beta_c$ the configuration subspace may have the property of full-replica-symmetry-breaking. We expect that the FA system in the kinetic cluster $C_1$ will also show property similar to the Gardner phase of structural glasses~\cite{Urbani-etal-2023,Liao-etal-2023,Berthier-etal-2016}. This is an interesting and challenging topic for further exploration, especially for finite-dimensional systems.

\section{$K$-core phase transition}
\label{sec:kpl}

Starting from an initial occupation configuration $\vec{\bm{c}}^{\,0}$ that happens to reside in certain kinetic cluster $C_\alpha$ of the graph $G$, the evolution trajectory of the FA kinetic system is then confined to this cluster and it will not jump to another different one. To determine which vertices can change occupation states in this cluster $C_\alpha$ and which are completely frozen, we can perform the simple $K$-core pruning process with $\vec{\bm{c}}^{\,0}$ as the initial condition: If a vertex $i$ has less than $K$ occupied nearest neighbors, then it is added to the set of unfrozen vertices and is deleted from the graph; and this is repeated on the vertices of the remaining graph until no further modification can be made. The vertices of the final residual graph are then either frozen to the occupied state or frozen to the empty state. It is straightforward to prove that the frozen occupied vertices must form a $K$-core within themselves. We denote by $f$ the fraction of vertices in this $K$-core ($f$ is also its relative size). The $K$-core relative size $f$ can be estimated by a simple mean-field theory for a finite-connectivity random graph $G$~\cite{Branco-1993,Pittel-etal-1996}. This is also an interesting topic of network science~\cite{Dorogovtsev-etal-2006,Zhao-Zhou-Liu-2013,Rizzo-2018,Bianconi-Dorogovtsev-2024b}.

\subsection{Regular random graph ensemble}

As a concrete demonstration, we consider the ensemble of regular random graphs with uniform degree $D$. The initial fraction of occupied vertices is $p$, and these vertices are drawn uniformly at random from the graph. A crucial empirical observation underlying this percolation theory is the following: If an extensive frozen $K$-core emerges in the graph $G$, the other complementary part of the graph will not be densely connected but will be a collection of (almost) loop-free tree components.

The fraction $f$ of frozen occupied vertices in the regular random graph is equal to the probability of a randomly chosen vertex $i$ being frozen to $c_i = 1$. Assuming statistical independence among the nearest neighbors of this focal vertex $i$, we can write down the following expression:
\begin{equation}
  \label{eq:kinetic_qi}
  f = p \sum\limits_{d=K}^{D} \frac{D!}{d! (D-d)!} \gamma^{d} (1 - \gamma)^{D-d}
  \; .
\end{equation}
Here $\gamma$ is the conditional probability of a nearest neighbor $j$ being frozen to the occupied state, given that the state of vertex $i$ is fixed to $c_i = 1$. The self-consistent equation for $\gamma$ is similar to Eq.(\ref{eq:kinetic_qi}),
\begin{equation}
  \gamma = p \sum\limits_{d=K-1}^{D-1} \frac{(D-1)!}{d! (D-1-d)!}
  \gamma^d (1 - \gamma)^{D-1-d} \; .
  \label{eq:gamma}
\end{equation}
This equation always has a trivial fixed point $\gamma = 0$, corresponding to the absence of frozen $K$-core ($f=0$). 

When $K = 2$, the stable fixed point $\gamma$ of Eq.~(\ref{eq:gamma}) starts to gradually deviate from zero at certain critical value $p_c$ of the initial occupation density $p$. Consequently, the fraction $f$ of frozen occupied vertices also gradually increases from zero as $p$ increases from $p_c$. A continuous $2$-core percolation transition occurs at $p = p_c$, similar to the conventional site percolation discussed in percolation theory~\cite{He-Liu-Wang-2009,Li-etal-2021}. After this transition, the frozen occupied vertices form an extensive $2$-core containing $N f$ vertices.

\begin{figure}[b]
  \centering
  \includegraphics[angle=270,width=0.7\linewidth]{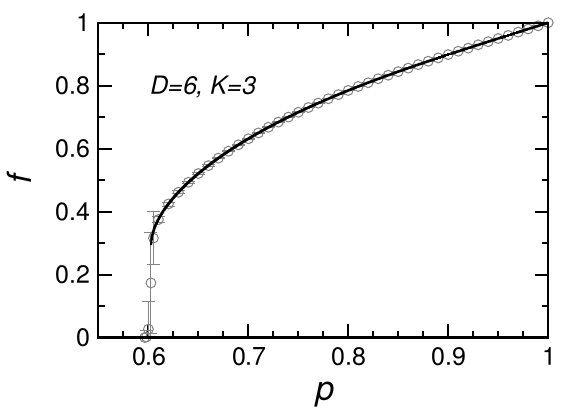}
  \caption{
    The relative size $f$ of the frozen  occupied $K$-core with $K=3$ versus the initial fraction $p$ of occupied vertices, for the regular random graph ensemble of degree $D=6$. Solid line is theoretical prediction (\ref{eq:kinetic_qi}). Circles and error bars (denoting standard deviation) are simulation results obtained on a single graph instance of size $N = 32768$~\cite{Wang-etal-2020}.
  }
  \label{fig:KineticRRD6}    
\end{figure}

When $K \geq 3$, however, the trivial fixed point $\gamma = 0$ is always locally stable, and another stable fixed point with $\gamma$ considerably larger than zero appears suddenly as $p$ exceeds a critical value $p_c$. The theory therefore predicts a sudden emergence of an extensive $K$-core with the relative size $f$ jumping from $f=0$ to a large value, see Fig.~\ref{fig:KineticRRD6} for an example~\cite{Wang-etal-2020}. This explosive emergence of a giant frozen $K$-core has been verified by computer simulations~\cite{Sellitto-etal-2005,Schwarz-Liu-Chayes-2006,Perrupato-Rizzo-2023}. 

The existence of a discontinuous $K$-core percolation phase transition for $K \geq 3$ means that, the maximally random occupation configurations $\vec{\bm{c}}^{\,0}$ with occupation density $p > p_c$ are highly likely to be kinetically jammed configurations. Extensively many vertices are unable to change their states under the kinetic rule of the FA system. A dynamical phase transition occurs at $p = p_c$ in the evolution trajectories of the FA system. Information about the initial configuration $\vec{\bm{c}}^{\,0}$ will be largely reserved even as time goes to infinity, if the fraction of occupied vertices in $\vec{\bm{c}}^{\,0}$ is larger than $p_c$, while this memory will be completely lost with time if $p$ is less then $p_c$.

\subsection{Periodic cubic lattice}

The $K$-core percolation in finite-dimensional regular lattices has also been extensively investigated in the literature~\cite{Adler-1991,Rizzo-2018,Sur-etal-1976,vanEnter-1987,Schonmann-1990,Schonmann-1992,Lin-Hu-1998,Branco-Silva-1999,Kurtsiefer-2003,Gregorio-etal-2004,Zhu-etal-2015}. Here we show some representative results obtained on the cubic lattice containing $N=L^3$ vertices, were $L$ is the side length of the lattice. Periodic conditions along each of the three dimensions are imposed, so every vertex has $D=6$ nearest neighbors. Such lattices are better models than random graphs for understanding real-world physical systems. 

\begin{figure}[t]
  \centering
  \subfigure[]{
    \includegraphics[angle=270,width=0.7\linewidth]{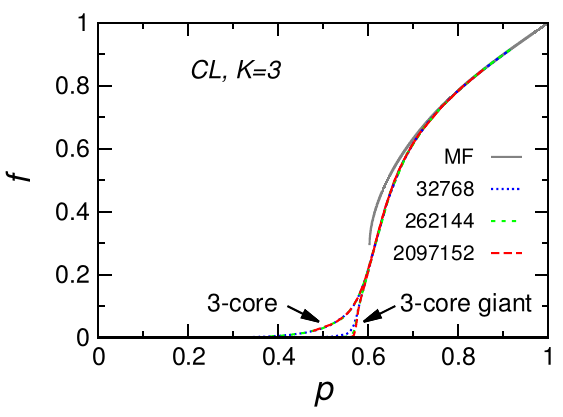}
    \label{fig:LC3coref}
  }
  \subfigure[]{
    \includegraphics[angle=270,width=0.7\linewidth]{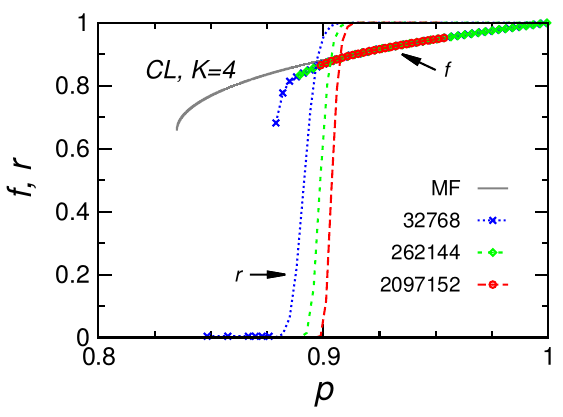}
    \label{fig:LC4coref}
  }
  \caption{
    $K$-core percolations on periodic cubic lattice with uniform vertex degree $D=6$, at $K=3$ (a) and $K=4$ (b). The mean relative size $f$ of $K$-core patterns at each value $p$ of initial occupation density is obtained by averaging over $20000$ independent samples. The relative size of the giant component of the $3$-core is also shown in (a).  The fraction $r$ of samples containing an extensive $4$-core is also shown in (b), and here the value of $f$ is averaged only over these non-trivial samples. The solid curves are the predictions of the mean-field (MF) formula (\ref{eq:kinetic_qi}). The total number of vertices are $N=32768$, $262144$ and $2097152$, corresponding to side lengths $L=32$, $64$ and $128$, respectively.
  }
  \label{fig:LCKcore}
\end{figure}

First, consider the case of $K=3$ (which is also representative of $K=1, 2$). The simulation results on the mean relative size $f$ of $3$-core are plotted in Fig.~\ref{fig:LC3coref}. We see that the curve $f(p)$ is a continuous and smooth function of the initial occupation density $p$ and it is almost independent of $N$ for side lengths $L \geq 32$. This means that the emergence of $3$-core itself is not a phase transition phenomenon. This smooth behavior of $f(p)$ is a consequence of the lattice geometric structure and is relatively easy to understand. As long as $p > 0$, many minimum cubes formed by four occupied vertices are dispersed in the cubic system and are contributing to the extensive size of the $3$-core. This is also the situation for $K=1$ and $K=2$. 

Although there is no phase transition of $3$-core emergence in the periodic cubic lattice system, there is a continuous phase transition concerning the emergence of an extensive connected component within the $3$-core [Fig.~\ref{fig:LC3coref}]. This continuous phase transition is resembling the emergence of an extensive giant component in conventional site percolation and it is amenable to finite-size scaling analysis~\cite{Zhu-etal-2015}. The phase transition point has been located at the critical value of $p_c \approx 0.5726$~\cite{Branco-Silva-1999}. We also see from Fig.~\ref{fig:LC3coref} that, when the initial occupation density $p$ exceeds $0.75$, the $3$-core relative size can again be well described by the mean field formula (\ref{eq:kinetic_qi}).

Similar continuous phase transitions also occur for $K=1$ and $K=2$. However, when $K\geq 4$, the $K$-core percolation on the periodic cubic lattice is a discontinuous process. Some example results obtained for $K=4$ are shown in Fig.~\ref{fig:LC4coref}. For a given system size $L$ (and number of vertices $N=L^3$), when the initial fraction $p$ of occupied vertices is smaller than certain critical value $p_c(L)$, there is high chance that the random initial configuration $\vec{\bm{c}}^{\,0}$ contains no $4$-core. When $p > p_c(L)$, with high probability there is an extensive $4$-core, and its relative size $f$ is close to the value predicted by the mean-field theory (\ref{eq:kinetic_qi}). The value of $p_c(L)$ for each $L$ is determined as the point of $p$ at which fifty percent of the independent samples $\bm{c}^{\,0}$ contain an extensive $4$-core.

Another significant feature of Fig.~\ref{fig:LC4coref} is that the threshold value $p_c(L)$ is not fixed but slowly increasing with the side length $L$. When the periodic cubic lattice becomes infinite, the threshold value will approach unity, $p_c(L=\infty)= 1$~\cite{vanEnter-1987,Schonmann-1990,Schonmann-1992}. The key to understand this asymptotic behavior is the shape irregularity of a local connected domain of empty vertices. Some of the occupied vertices at the boundary of such an irregular empty domain are unstable and will join this domain. If the enlarged empty domain during this erosion process always deviates significantly from being a cubical shape, it will keep growing. Such rare self-growing empty domains are deemed to be present in an infinite cubic lattice~\cite{vanEnter-1987,Gregorio-etal-2004}.

This type of extreme cascading behavior may be absent in the case of heterogeneous $\{k_i\}$--core dynamics, with a finite fraction of the threshold values $k_i$ being $k_i \leq 2$ and with these low-threshold vertices $i$ being randomly distributed in the cubic lattice. More thorough simulation studies are needed to clarify this point~\cite{Branco-1993,Cellai-etal-2011,Baxter-etal-2011}.  Polydispersity of the threshold values $\{k_i\}$ may be corresponding to polydispersity of particle size or shape in structural glass systems.

We close this section by emphasizing that the kinetic $K$-core patterns studied here are corresponding to completely random and uncorrelated initial occupation configurations $\vec{\bm{c}}^{\,0}$. For a given relative size $f$, these kinetic $K$-core patterns may far from being the equilibrium (typical) $K$-core patterns to be discussed in the next section.

\section{Exploring partially frozen kinetic clusters}
\label{sec:kfr}

As we mentioned earlier (Sec.~\ref{subsec:Calpha}), a partially frozen kinetic cluster $C_\alpha$ of the FA system is completely determined by its set of frozen occupied vertices, and these frozen vertices form a jammed $K$-core pattern in the graph $G$. The total number $\mathcal{N}$ of kinetic clusters is huge, and the jammed $K$-cores obtained through randomized initial conditions (Sec.~\ref{sec:kpl}) are unlikely to be typical $K$-core patterns for the given relative size $f$. In this section we describe an equilibrium $K$-core statistical physics problem that is quite helpful for exploring the statistical properties of typical jammed $K$-core patterns.

\subsection{Statistical mechanics of $K$-core jamming}
\label{subsec:ekcj}

The total number $\mathcal{N}$ of possible kinetic clusters for a given graph $G$ is obtained by counting the total number of jammed (frozen) $K$-cores that can be constructed for this graph. Let us denote by $f$ the fraction of vertices in such a frozen occupied $K$-core, and denote by $s(f)$ the entropy density of $K$-cores containing  $N f$ frozen occupied vertices. Then we have
\begin{equation}
  \mathcal{N} \ = \ 1 + 
  \int _{f_{\textrm{min}}}^1 \textrm{d} f \, e^{N s(f) } \; ,
  \label{eq:clusternumber}
\end{equation}
where $f_{\textrm{min}}$ is the minimum fraction of occupied vertices needed to sustain a non-vanishing $K$-core.

Computing the entropy density $s(f)$ and determining the minimum occupation fraction $f_{\textrm{min}}$ are nontrivial tasks. Recently we introduced an equilibrium $K$-core model to study the network defensive alliance problem and the network engagement problem~\cite{Xu-etal-2018,Wang-etal-2020}. Since partially frozen kinetic FA clusters are essentially jammed $K$-core patterns, this same model is naturally also applicable here. In this setting, each vertex $i$ has an aggregated state $a_i \in \{0, 1\}$: $a_i = 0$ means that it is either initially empty or is unfrozen, and $a_i = 1$ means that it is always frozen to be occupied. We assign energy values $0$ and $1$ to these two aggregated states $a_i=0$ and $a_i=1$, respectively.

The equilibrium partition function for this system, $Z^{\textrm{jam}}(\beta)$, is then written as~\cite{Xu-etal-2018,Wang-etal-2020}
\begin{equation}
  Z^{\textrm{jam}}( \beta )  =  \sum\limits_{\{a_i \}\neq \vec{\bm{0}}}
  \prod\limits_{i=1}^{N} \biggl[ \delta_{a_i}^0 + \delta_{a_i}^1
    e^{-\beta} I\Bigl( \sum_{j\in \partial i} a_j \geq K \Bigr) \biggr] \;
  \; ,
  \label{eq:Zkcore}
\end{equation}
where the subscript $\{a_i\} \neq \vec{\bm{0}}$ means that the all-empty pattern $\vec{\bm{0}}$ is excluded from the summation. Here the inverse temperature $\beta$ controls the size of the $K$-core. The indicator function $I\bigl(\sum_{j\in \partial i} a_j \geq  K \bigr) \in \{0, 1\}$ means that if a vertex $i$ belongs to the $K$-core ($a_i = 1$), then at least $K$ nearest neighbors of this vertex must also belong to the $K$-core. For any valid $K$-core pattern $\vec{\bm{a}}=(a_1, \ldots, a_N)$, the equilibrium Boltzmann distribution associated with the partition function $Z^{\textrm{jam}}(\beta)$ is simply
\begin{equation}
  P_B^{\textrm{jam}}( \vec{\bm{a}} ) \, = \,
  \frac{1}{Z^{\textrm{jam}}(\beta) } \prod\limits_{i=1}^{N}
  e^{- \beta a_i } \; .
  \label{eq:PBjam}
\end{equation}
We denote by $f$ the fraction (density) of occupied vertices in a $K$-core pattern, namely $f = (1/N) \sum_i a_i$. The mean relative size of the equilibrium $K$-cores at a given inverse temperature is then the average of $f$ among all the valid $K$-core patterns sampled following Eq.~(\ref{eq:PBjam}).

When $\beta$ becomes sufficiently high, only the $K$-core patterns with minimum relative size $f_{\textrm{min}}$ will be relevant, and the partition function is $Z^{\textrm{jam}} \approx e^{- \beta N f_{\textrm{min}}}$. A minimum $K$-core pattern is extremely sensitive to external perturbations, since deleting any of its member vertices will make the whole $K$-core completely collapse. This also indicates that a minimum $K$-core pattern must be very difficult to access in the FA kinetic system (the initial configuration $\vec{\bm{c}}^{\,0}$ must be chosen extremely carefully).

\subsection{Entropy density, and minimum $K$-core size}
\label{subsec:kcoreentropy}

\begin{figure}[t]
  \centering
  \includegraphics[angle=270,width=0.7\linewidth]{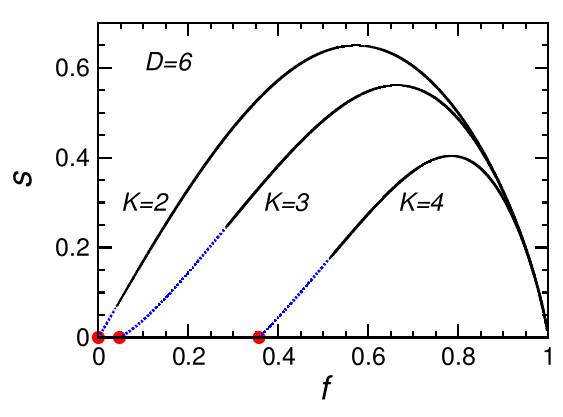}
  \caption{
    The entropy density $s$ versus the relative size $f$ of the $K$-core ($K=2,3,4$), for the regular random graph ensemble with degree $D=6$. The convex branch of each $s(f)$ curve is highlighted with the dotted line segment. Filled circles mark the minimum relative size $f_{\textrm{min}}$. Figure adapted from Ref.~\cite{Wang-etal-2020}, with kind permission of Europhysics Letters.
  }
  \label{fig:RRD6:a}
\end{figure}

We applied the mean field replica-symmetric cavity method to the model (\ref{eq:Zkcore}), assuming the underlying graph being regular random with uniform degree $D$~\cite{Xu-etal-2018,Wang-etal-2020}. Figure~\ref{fig:RRD6:a} demonstrates some representative theoretical results concerning the entropy density.

For each fixed value of $K$, the entropy density $s(f)$ has a single maximum at certain intermediate value $f^*$. This implies that the most abundant $K$-core patterns have certain intermediate relative size $f^*$, and the total (exponential) number of kinetic clusters is, to leading order, $\mathcal{N} \approx e^{N s(f^*)}$. For $K=3$ and $D=6$, we have $f^* \approx 0.663$ and $s(f^*) \approx 0.561$. Notice that the entropy density of kinetic clusters, $(1/N) \ln \mathcal{N}$, is much lower than the entropy density $\ln 2$ ($\approx 0.693$) of microscopic configurations, as it should be.
The entropy of a conventional statistical physics system is a concave function of energy. However, for the $K$-core jamming system (\ref{eq:Zkcore}), we find that its entropy density $s(f)$ is concave only if $f$ is higher than certain critical value $f_{\textrm{x}}$ (the inflection point~\cite{Xu-etal-2018}). When the $K$-core relative size is less than $f_{\textrm{x}}$, the entropy density becomes a convex function, and the minimum value $f = f_{\textrm{min}}$ is located at the end of this convex part.  For $K=3$ and $D=6$ the predicted minimum is $f_{\textrm{min}} \approx 0.047$ and the inflection point is $f_{\textrm{x}} \approx 0.282$. Notice that the critical relative size of $K$-cores obtained through random deletion of occupied vertices, $f = 0.294$ (Fig.~\ref{fig:KineticRRD6}), is considerably higher than $f_{\textrm{x}}$.

This convex property of $s(f)$ means that partially frozen kinetic clusters whose  jammed $K$-core relative sizes $f \in (f_{\textrm{min}}, f_{\textrm{x}})$ are invisible in the conventional thermodynamic sense. If we perform a simulated annealing dynamical process to sample jammed $K$-core patterns with $\beta$ increasing at certain rate, it is relatively easy to get $K$-core patterns with relative sizes $\geq f_{\textrm{x}}$, but the minimum-sized patterns with $f \approx f_{\textrm{min}}$ can not be achieved. The sampled jammed $K$-cores instead often exhibit a sudden drop in their relative size $f$, from $f_{\textrm{x}}$ to an intermediate value between $f_{\textrm{x}}$ and $f_{\textrm{min}}$, and then keep fluctuating slightly around this value~\cite{Xu-etal-2018}.

Similar to the kinetic procedure of Sec.~\ref{sec:kpl}, we can also get frozen $K$-core of occupied vertices by starting from an initial occupation configuration $\vec{\bm{c}}^{\,0}$. Depending on the particular protocol of sampling $\vec{\bm{c}}^{\,0}$, $K$-core jammed patterns with different relative sizes $f$ may be reached. In other words, we may get a ``jamming line'' which ends at $f=1$ and possibly begins at $f = f_{\textrm{x}}$. We are not yet certain whether or not this jamming line might have some qualitative similarity with the jamming line and the jamming plane of sphere packings~\cite{Jin-Yoshino-2021,Wilken-etal-2021,Ozawa-etal-2017,Woodcock-2013,Li-etal-2010b,Jin-Makse-2010,Pan-etal-2023}. Future more systematic investigations may resolve this issue.

\subsection{Extreme vulnerability of $K$-core}
\label{subsec:ultra}

Given a $K$-core jamming pattern $\vec{\bm{a}} = (a_1, \ldots, a_N)$ of occupied vertices, we say a vertex $i$ is a ``break point'' if it belongs to this $K$-core ($a_i = 1$) and its flipping ($a_i \rightarrow 0$) will cause the $K$-core to break down completely~\cite{Wang-etal-2020}. We denote by $\phi$ the fraction of break-point vertices. If this fraction is positive, then the $K$-core pattern $\vec{\bm{a}}$ must be extremely vulnerable to single-vertex perturbations. Obviously we have $\phi( \vec{\bm{a}} ) \leq f( \vec{\bm{a}} )$ for any $K$-core pattern $\vec{\bm{a}}$. In the limiting case of $\vec{\bm{a}}$ being a minimum $K$-core, then $\phi = f = f_{\textrm{min}}$ (all the members are break points).

When the occupation density $f$ of the equilibrium $K$-core patterns sampled according to Eq.~(\ref{eq:PBjam}) is sufficiently high, the $K$-core patterns are quite robust to single-vertex perturbations. Deleting any member vertex from the $K$-core will only cause small shrinkage in its size, and the fraction of break points is $\phi = 0$. However, when $f$ is lower than certain threshold value, the equilibrium $K$-core patterns become extremely vulnerable to single-vertex perturbations, and the fraction $\phi$ of break points becomes positive. We have quantitatively studied this phase transition to extreme vulnerability by a replica-symmetric mean-field theory and by computer simulations~\cite{Wang-etal-2020}.

\begin{figure}[t]
  \centering
  \subfigure[]{
    \includegraphics[angle=270,width=0.7\linewidth]{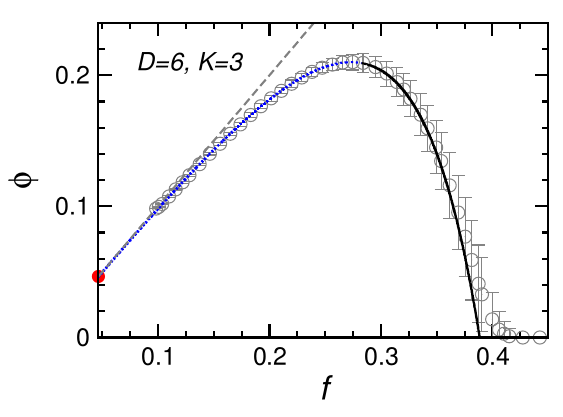}
    \label{fig:RRD6phi}
  }
  \subfigure[]{
    \includegraphics[angle=270,width=0.7\linewidth]{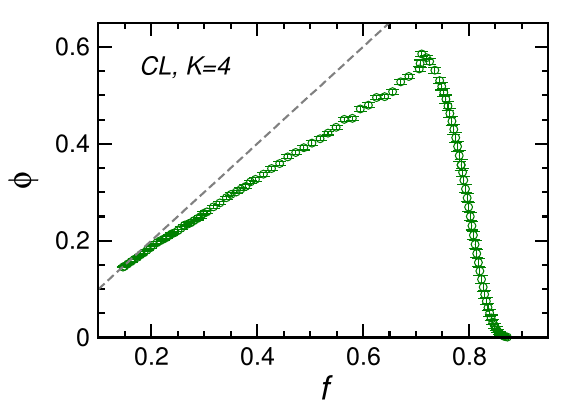}
    \label{fig:CL4corephi}
  }
  \caption{
    Fraction $\phi$ of break-point vertices versus the relative size $f$ of $K$-core. Symbols are simulation results obtained on a single graph instance of size $N = 32768$, and error bars denote standard deviations. Dashed line denotes the upper bound $\phi = f$.  (a) Regular random graph ensemble of degree $D=6$, with $K=3$. Solid and dotted line segments are theoretical predictions within the concave and convex region of the entropy density, respectively. Filled circle marks the minimum relative size $f_{\textrm{min}}$. (b) Periodic cubic lattice of side length $L=32$ and degree $D=6$, with $K=4$.  Figure (a) adapted from Ref.~\cite{Wang-etal-2020}, with kind permission of Europhysics Letters.
  }
  \label{fig:phi}
\end{figure}

Some representative results obtained on the regular random graph ensemble of degree $D=6$ and threshold value $K=3$ are shown in Fig.~\ref{fig:RRD6phi}. The order parameter $\phi$ starts to be positive as the relative size $f$ of $K$-core is lowered to the critical value $f_{\textrm{wt}}=0.388$, which is referred to as the weak tipping point in Ref.~\cite{Wang-etal-2020}. The ratio of $\phi$ versus $f$ then increases continuously as $f$ is further decreased, and this ratio approaches unity as the minimum relative size $f_{\textrm{min}}$ is approached from above.

It is interesting to notice that the weak tipping point $f_{\textrm{wt}}$ is distinctively higher than the entropy inflection point $f_{\textrm{x}}$ [Fig.~\ref{fig:RRD6phi}]. This may be a common phenomenon for the equilibrium $K$-core problem on random graph ensembles. 

The fact that a finite fraction $f$ of vertices are break points means that the partially frozen kinetic clusters $C_\alpha$ of the FA system, with relatively low density of frozen occupied vertices ($f < f_{\textrm{wt}}$), are staying at the surface of the fully unfrozen kinetic cluster $C_1$. Just a single ``forced'' flip of one break-point vertex is enough to melt the whole cluster $C_\alpha$ and make it merge with a sub-region of the cluster $C_1$.  This fact also implies the existence of long and complicated paths within $C_1$ between the all-empty $\vec{\bm{0}}$ and the configurations at the interface to $C_\alpha$. The surface of cluster $C_1$ in the high-dimensional configuration space must have very complex structures.

The vulnerability of equilibrium $K$-core patterns of structured graph instances were also investigated in Ref.~\cite{Wang-etal-2020}. It appears that extreme vulnerability is a general property of $K$-core patterns of relatively low occupation density $f$. For example, on the three-dimensional cubic lattice with periodic boundary condition, whose vertex degree is $D=6$ and size is $N=32768$ (side length $L=32$), the equilibrium $K$-core patterns with $K=4$ starts to have positive $\phi$ at a high value of $f \approx 0.83$ [Fig.~\ref{fig:CL4corephi}]. The clear cusp close to the maximum value of $\phi$ in the curve $\phi(f)$ of Fig.~\ref{fig:CL4corephi} may indicate some phase separation behavior.

According to the simulation results of Fig.~\ref{fig:CL4corephi}, $\phi$ approaches $f$ as the relative size of equilibrium $4$-core patterns is reduced to $f\approx 0.15$, indicating that the minimum value of relative size must be $f_{\textrm{min}} \leq 0.15$. We notice that this minimum value is much lower than the typical relative sizes of random kinetic $4$-core patterns [Fig.~\ref{fig:LC4coref}].

\section{Conclusion and outlook}
\label{sec:conc}

In this paper, we reviewed some recent statistical physics results on two combinatorial optimization challenges associated with the $K$-core collective structure of random graphs, the $K$-core attack problem and the equilibrium $K$-core problem. The emphasize of this \emph{Perspective} has been on the link between the $K$-core optimization problems and the thermodynamical properties of the Fredrickson-Andersen kinetically constrained spin model. In our opinion, the $K$-core attack problem is a perfect model for studying the thermodynamic phase transitions of the FA kinetic system, and in return, the FA kinetic system can be employed as an efficient local algorithm to sample equilibrium configurations for the globally constrained $K$-core attack problem (Section~\ref{sec:global}). We hope our work will stimulate more future work on these and other $K$-core optimization problems.

The statistical physics properties of the $K$-core attack problem and the equilibrium $K$-core problem on finite-dimensional lattices are not yet clear. Systematic numerical and theoretical efforts along this direction, especially concerning the spin glass phase transitions, will be highly desirable in the near future. They will be indispensable for us to fully understand the thermodynamic properties of the finite-dimensional FA kinetic systems.

The global constraint of $K$-core absence makes the $K$-core attack problem difficult to tackle theoretically. Although some progresses have been made in recent years, the spin glass phase transition issues are largely unexplored, even for random graph ensembles. To explore the rich structure of the low-temperature (high-occupation density) energy landscape, one particularly interesting non-equilibrium issue is to study the quenched dynamics of the $K$-core attack model at $\beta = \infty$ (see Ref.~\cite{Folena-etal-2020} for a related recent work).

We have discussed unjammed maximum packing ($\rho_{\textrm{max}}$) and jammed minimum packing ($f_{\textrm{min}}$) in the FA kinetic system. Similar quantities are also investigated in the field of structural glasses. Structural glasses and amorphous solid are of course more complicated than the FA model and other kinetically constrained spin models. Particles in a structural glass system have translational degrees of freedom, they will diffuse with time and change their interacting partners.  They may also have different sizes and different shapes, and there are friction forces between the contacting particles (see, for example, Refs.~\cite{Liu-Tong-etal-2020,Chen-etal-2023,Pan-etal-2023,Liao-etal-2023}  for a glimpse of recent simulation and experimental results). But we feel it may be plausible to expect that the kinetic constraints encountered by the particles during the translational and rotational motions will also bring certain global constraint to the microscopic spatial configurations of the structural glass system, similar to the $K$-core-absence constraint induced by the kinetic rules of the FA model. If this is the case, then the $K$-core attack problem will also be relevant for structural glass studies. 

\begin{acknowledgments}
  The author thank Luan Cheng, Yang-Yu Liu, Shao-Meng Qin, David Saad, Shang-Nan Wang, Yi-Zhi Xu, Chi Ho Yeung, Ying Zeng, Pan Zhang, Jin-Hua Zhao, and Jianwen Zhou for collaborations on the $K$-core optimization problems. The valuable comments of Qinyi Liao on an earlier version of this manuscript is warmly acknowledged. The author also benefited from discussions with Xinyi Fan on the $K$-core preoclation in cubic lattices. The following funding supports are acknowledged: National Natural Science Foundation of China Grants No.~12247104, 12047503.
\end{acknowledgments}

\end{document}